\documentclass[
  twocolumn,
  prl,
  amsmath,
  amssymb,
  superscriptaddress,
  floatfix
]{revtex4}

\usepackage{bm}
\usepackage{graphicx}

\begin{document}

\title{Quantum criticality and minimal conductivity in graphene with long-range disorder}

\author{P.~M.~Ostrovsky}
\affiliation{
 Institut f\"ur Nanotechnologie, Forschungszentrum Karlsruhe,
 76021 Karlsruhe, Germany
}
\affiliation{
 L.~D.~Landau Institute for Theoretical Physics RAS,
 119334 Moscow, Russia
}

\author{I.~V.~Gornyi}
\altaffiliation{
 Also at A.F.~Ioffe Physico-Technical Institute,
 194021 St.~Petersburg, Russia.
}
\affiliation{
 Institut f\"ur Nanotechnologie, Forschungszentrum Karlsruhe,
 76021 Karlsruhe, Germany
}

\author{A.~D.~Mirlin}
\altaffiliation{
 Also at Petersburg Nuclear Physics Institute,
 188350 St.~Petersburg, Russia.
}
\affiliation{
 Institut f\"ur Nanotechnologie, Forschungszentrum Karlsruhe,
 76021 Karlsruhe, Germany
}
\affiliation{
 Inst. f\"ur Theorie der kondensierten Materie,
 Universit\"at Karlsruhe, 76128 Karlsruhe, Germany
}

\date{\today}

\begin{abstract}
We consider the conductivity $\sigma_{xx}$ of graphene with negligible
intervalley scattering at half filling. We derive the effective field theory,
which, for the case of a potential disorder, is a symplectic-class
$\sigma$-model including a topological term with $\theta=\pi$. As a
consequence, the system is at a quantum critical point with a universal value
of the conductivity of the order of $e^2/h$. When the effective time reversal
symmetry is broken, the symmetry class becomes unitary, and $\sigma_{xx}$ 
acquires the value characteristic for the quantum Hall transition. \\[-0.7cm]
\end{abstract}

\maketitle

Recent breakthrough in graphene fabrication \cite{Novoselov04} and subsequent
transport experiments \cite{Novoselov05} revealed remarkable electronic
properties of this material. One of the most striking experimental observation
is the minimal conductivity $\sigma_{xx}$ of order $e^2/h$ observed in undoped
samples and staying almost constant in a wide range of temperatures $T$ from
300 K down to $\sim 1K$. This behavior should be contrasted to well-established
results on the conductivity of two-dimensional (2D) systems where localization
effects drive the system into insulating state at low $T$ \cite{McCann, AEA}.
An apparently $T$-independent value of $\sigma_{xx} \sim e^2/h$ suggests that
the system is close to a quantum critical point and calls for a theoretical
explanation.

One particular class of randomness when this scenario is realized, namely, the
chiral disorder, was analyzed in detail in Ref.\ \cite{OurPRB} (see also Ref.\
\cite{Ryu}). It was shown that, if one of the chiral symmetries of clean
graphene is preserved by disorder, the conductivity at half filling is not
affected by localization and is equal to $(4/\pi)e^2/h$ (up to small
corrections). While various types of randomness in graphene (in particular,
dislocations, ripples, or strong point-like defects)  do belong to the chiral
type, the experimentally observed value of $\sigma_{xx}$ is larger by a factor
$\sim 3$, suggesting a different type of criticality. In this paper we consider
another broad class of randomness in graphene --- long-range disorder. This
case has a particular experimental relevance if the conductivity is dominated
by charged impurities; the ripples \cite{Morpurgo06,Meyer07} belong to this
class of randomness as well. Numerical simulations of graphene with long-range
random potential \cite{Nomura06, Beenakker_numerics} provide an evidence in
favor of a scale-invariant conductivity.

The low-energy electron spectrum of clean graphene split into two degenerate
valleys. The characteristic feature of the long-range disorder is the absence
of valley mixing due to the lack of scattering with large momentum transfer.
This allows us to describe the system in terms of a single-valley Dirac
Hamiltonian with disorder,
\begin{equation}
 H
  = v_0 \bm{\sigma}\mathbf{k} + \sigma_\mu V_\mu(\mathbf{r}).
 \label{ham}
\end{equation}
Here $v_0 \simeq 10^8$ cm/s is the Fermi velocity. The four Pauli matrices
$\sigma_\mu$ (with $\sigma_0 = 1$) operate in the space of two-component
spinors reflecting the sublattice structure of the honeycomb lattice,
$\bm{\sigma} = \{ \sigma_1, \sigma_2 \}$, and disorder includes random scalar
($V_0$) and vector ($V_{1,2}$) potentials and random mass ($V_3$). The
Hamiltonian (\ref{ham}) was considered in Ref. \cite{Ludwig} as a model for
quantum Hall transition.

To derive the field theory, we introduce a vector superfield $\psi$ with
$2\times 2 \times 2 = 8$ components: the matrix structure of $H$ in the
sublattice space is complemented by the boson--fermion (BF) and the
retarded-advanced (RA) structures. Assuming for simplicity Gaussian
$\delta$-correlated disorder distribution, we get the action
\begin{equation}
 S[\psi]
  = \int d^2 r \biggl[
      -i \bar\psi \bigl(
        \hat\varepsilon + i v_0 \bm{\sigma} \nabla
      \bigr) \psi + \frac{1}{2} \langle V_\mu^2\rangle (\bar\psi
      \sigma_\mu \psi)^2
    \biggr],
 \label{action}
\end{equation}
with $\hat\varepsilon = \varepsilon + i0\Lambda$ where $\Lambda =
\mathop{\mathrm{diag}} \{1, -1\}_{RA}$. Assuming the isotropy, the disorder is
described by three couplings $\alpha_0 = \langle V_0^2 \rangle / 2 \pi v_0^2$,
$\alpha_\perp = \langle V_1^2 + V_2^2 \rangle / 2 \pi v_0^2$, and $\alpha_z =
\langle V_3^2 \rangle / 2 \pi v_0^2$. On short (ballistic) scales the
parameters of (\ref{action}) are renormalized \cite{Ludwig, Nersesyan, AEA,
OurPRB}; the effective theory on longer scales is the non-linear sigma model
\cite{Efetov}.

The clean single-valley Hamiltonian (\ref{ham}) obeys the effective
time-reversal (TR) invariance $H=\sigma_2 H^*\sigma_2$. This symmetry (denoted
as $T_\perp$ in Ref.\ \cite{OurPRB}) is not the physical TR symmetry: the
latter interchanges the nodes and is of no significance in the absence of
inter-node scattering.  If the only disorder is random scalar potential, the TR
invariance is not broken and the system falls into the symplectic symmetry
class (AII) \cite{Ludwig, McCann, AEA}. The standard realization of such
symmetry is a system with spin-orbit coupling; in the present context the role
of spin is played by the sublattice space.

We start with a more generic case of the unitary symmetry (class A). The TR
invariance is broken as soon as a (either random or non-random) mass or vector
potential is included, in addition to the scalar potential. We find it
instructive to present the derivation for a system with a non-zero mass term $m
\sigma_3$. Decoupling the $\psi^4$ term by a supermatrix field $Q$ and
integrating out $\psi$, we get
\begin{equation}
 S[Q]
  = \mathop{\mathrm{Str}} \left[ -\frac{\gamma^2 Q^2}{4 \pi v_0^2 \alpha_0}
      +\ln \Bigl(
        \varepsilon - m \sigma_3 + i v_0 \bm{\sigma} \nabla + i \gamma Q
      \Bigr)
    \right],
 \label{action_log}
\end{equation}
where {\rm Str} includes the matrix supertrace and the spatial integration,
$\gamma=1/2\tau$, and $\tau$ is the mean free time.

The saddle-point approximation \cite{footnote_critical} reduces the set of $Q$
to the conventional manifold of the unitary $\sigma$-model; the relevant $Q$'s
are $4 \times 4$ supermatrices operating in RA and BF spaces and satisfying the
constraint $Q^2=1$. The low-energy modes describe slow spatial variation of $Q$
on this manifold, and the effective theory is the result of the gradient
expansion of the action (\ref{action_log}) in these modes. As we show, this
expansion is highly non-trivial due to anomalies in the theory of Dirac
fermions \cite{Haldane}, which induce a topological contribution to the
$\sigma$-model \cite{Bocquet}.

We first perform the gradient expansion of the real part of the action
(\ref{action_log})
\begin{equation}
 S_1[Q]
  = (1/2) \mathop{\mathrm{Str}} \ln \Bigl(
      G_+^{-1} G_-^{-1} + v_0 \gamma \bm{\sigma} \nabla Q
    \Bigr).
 \label{action_real}
\end{equation}
Here the matrix Green functions defined as $G_{\pm}^{-1} = \varepsilon - m
\sigma_3 + i \bm{\sigma} \nabla \pm i \gamma \Lambda$ are diagonal in RA space
with retarded and advanced Green functions as their elements, $G_+ =
\mathop{\mathrm{diag}} \{G^R, G^A\}$, $G_- = \mathop{\mathrm{diag}} \{G^A,
G^R\}$. Expanding Eq.\ (\ref{action_real}) to the second order in $\nabla Q$
and using the identity $G^R - G^A = -2 i \gamma G^R G^A$, we get the familiar
gradient term,
\begin{equation}
 S_1[Q]
  = (\sigma_{xx}/8) \mathop{\mathrm{Str}} \bigl( \nabla Q \bigr)^2.
\end{equation}
The factor $\sigma_{xx}$ in Eq.\ (\ref{action_real}) is identified as the
dimensionless (in units $e^2/h$) longitudinal conductivity given by
\begin{multline}
 \sigma_{xx}
  = -(v_0^2/2) \mathop{\mathrm{Tr}} \Bigl[
      \bigl( G^R - G^A \bigr) \sigma_x
      \bigl( G^R - G^A \bigr) \sigma_x
    \Bigr] \\
  = (1/2\pi) \left[
      1 + f(\varepsilon, m) (\varepsilon^2 + \gamma^2 - m^2)/2 \gamma
    \right],
 \label{sigma_xx}
\end{multline}
where we introduced  the notation
$$
 f(x, y)
  = x^{-1} \{
      \arctan [(x + y)/\gamma] + \arctan [(x - y)/\gamma]\}.
$$

The calculation of the imaginary part $iS_2[Q]$ is much more subtle. We use the
representation $Q = T^{-1} \Lambda T$ and cycle the matrices under the
supertrace. The resulting expression depends on the vector $\mathbf{u} = T
\nabla T^{-1}$,
$$
 i S_2[Q]
  = (1/2) \mathop{\mathrm{Str}} \Bigl[
      \ln \bigl(
        G_+^{-1} + i v_0 \bm{\sigma}\mathbf{u}
      \bigr)
      -\ln \bigl(
        G_-^{-1} + i v_0 \bm{\sigma}\mathbf{u}
      \bigr)
    \Bigr].
$$
The permutation of matrices leading to this formula is equivalent to a rotation
of fermion fields, $\psi \mapsto T \psi$, in Eq.\ (\ref{action}). This is not
an innocent procedure in view of quantum anomaly \cite{Fujikawa80}. However,
such anomalous contributions from the two logarithms cancel in $i S_2[Q]$. We
proceed with expanding $i S_2[Q]$ in powers of $\mathbf{u}$. The first two
terms of this expansion are
\begin{gather}
 i S_2^{(1)}\!
  = (i v_0/2) \mathop{\mathrm{Str}} \bigl[
      \Lambda \bigl( G^R - G^A \bigr) \bm{\sigma} \mathbf{u}
    \bigr]
  \equiv \pi \mathop{\mathrm{Str}} \bigl(
      \Lambda \mathbf{J} \mathbf{u}
    \bigr),  \label{action_img_1} \\
\! i S_2^{(2)}\!
  = \frac{\sigma_{xy}^I\epsilon_{\alpha\beta}}{2}  \mathop{\mathrm{Str}}
    \bigl(
      u_\alpha \Lambda u_\beta
    \bigr)\!
  = \frac{\sigma_{xy}^I}{4} \mathop{\mathrm{Str}} \bigl(
      Q \nabla_x Q \nabla_y Q
    \bigr).\label{action_theta_2}
\end{gather}
The factors in Eqs.\ (\ref{action_img_1}), (\ref{action_theta_2}) are the
current spectral density $\mathbf{J}(\mathbf{r})$ and the classical part of
Hall conductivity \cite{Pruisken}
\begin{multline}
 \sigma_{xy}^I
  = (v_0^2 / 2) \mathop{\mathrm{Tr}} \Bigl[
      \bigl( G^R + G^A \bigr) \sigma_x
      \bigl( G^R - G^A \bigr) \sigma_y
    \Bigr] \\
  = -(m / 2 \pi)\; f(\varepsilon, m).
\end{multline}
The net current, and hence the linear term (\ref{action_img_1}), is absent in
the bulk of the system. It is incorrect, however, to drop this term. The
contribution $i S_2^{(1)}$ accounts for the edge current and gives the quantum
part of the Hall conductivity \cite{Pruisken}. Prior to considering it, we have
to establish boundary conditions (BC) for the Hamiltonian (\ref{ham}).

Generically, BC in realistic graphene mix states from the two valleys of the
spectrum. We can stay, however, within the single-valley model and assume an
infinite mass $M \to \infty$ at the boundary of the sample
\cite{BerryMondragon}. Localization effects  described by the $\sigma$-model
occur in the bulk and hence are insensitive to particular BC. We thus assume
that $m(\mathbf{r})$ changes from a constant value $m$ inside the sample to
another, large value $M$ outside it. The gradient of mass is not zero near the
edge only. We further assume that the mass variation is slow on the scale of
the electron mean free path but fast compared to $\sigma$-model length scales.
This allows us to perform an expansion of the Green functions in Eq.\
(\ref{action_img_1}) in $\nabla m$. With the help of identity $[\mathbf{r}, G]
= i v_0 G \bm{\sigma} G$, we obtain
\begin{multline}
 J_\alpha(\mathbf{r})
  = - (v_0^2 / 2 \pi)\; \nabla_\beta m\; \mathop{\mathrm{tr}} \Bigl[
      \sigma_\alpha G^R \sigma_3 G^R \sigma_\beta G^R \\
      -\sigma_\alpha G^A \sigma_3 G^A \sigma_\beta G^A
    \Bigr]_{\mathbf{r}, \mathbf{r}}
  = \frac{\epsilon_{\alpha\beta}}{2 \pi}\;
    \frac{\partial \sigma_{xy}^{II}}{\partial m}\; \nabla_\beta m.
\label{jalpha}
\end{multline}
The emerged trace is a mass derivative of the quantum part $\sigma_{xy}^{II}$
of Hall conductivity \cite{Pruisken}; its direct calculation and subsequent
integration with respect to $m$ yields
\begin{equation}
 \sigma_{xy}^{II}
  = - (m /2 \pi)\; f(m, \varepsilon).
\end{equation}
Substituting (\ref{jalpha}) in (\ref{action_img_1}) and integrating over the
boundary strip, we express the term (\ref{action_img_1}) as an integral along
the edge and then apply the Stokes theorem:
\begin{multline}
 i S_2^{(1)}
  = \frac{\epsilon_{\alpha\beta}}{2} \Bigl[
      \sigma_{xy}^{II}(m) - \sigma_{xy}^{II}(M)
    \Bigr] \mathop{\mathrm{Str}} \bigl( \Lambda \nabla_\alpha u_\beta \bigr) \\
  = \frac{1}{4} \biggl(
      \sigma_{xy}^{II} + \frac{\mathop{\mathrm{sgn}} M}{2}
    \biggr) \mathop{\mathrm{Str}} \bigl(
      Q \nabla_x Q \nabla_y Q
    \bigr).
  \label{action_theta_1}
\end{multline}
To derive the last expression, we have used the identity
$\epsilon_{\alpha\beta} \nabla_\alpha u_\beta = \epsilon_{\alpha\beta} u_\beta
u_\alpha$ and the value of $\sigma_{xy}^{II}$ in the limit of infinitely large
$M$. The same result is obtained if one uses alternative BC introducing the
second node with the large mass $M$. In that case, $(1/2)\mathop{\mathrm{sgn}}
M$ will enter the action as the contribution of the second node to
$\sigma_{xy}$ (cf.\ Refs.\ \cite{Ludwig, Haldane}). Both contributions to
$iS_2[Q]$, Eqs.\ (\ref{action_theta_2}) and (\ref{action_theta_1}), contain the
functional $ \mathop{\mathrm{Str}} ( Q \nabla_x Q \nabla_y Q) \equiv 16 i \pi
N[Q]$ that is a well-known topological invariant on the $\sigma$-model manifold
\cite{Pruisken}; its possible values are integer multiples of $16 \pi i$. The
imaginary part of the action is defined up to a multiple of $2 \pi i$. Thus the
sign of $M$ is irrelevant, as expected: the bulk theory should not be sensitive
to BC.

\begin{figure}
\centerline{\includegraphics[width=0.8\columnwidth]{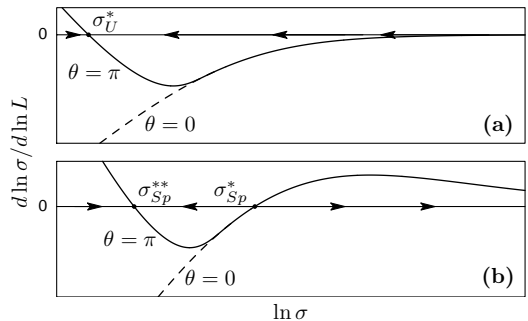}}
\caption{Schematic scaling functions for (a) unitary and (b) symplectic
universality class with topological term $\theta = \pi$. The ordinary case
$\theta = 0$ is shown by dashed lines.}
\label{Fig:beta}
\vspace*{-0.5cm}
\end{figure}

Collecting all the contributions, we get the $\sigma$-model action for the
single-node Dirac fermions with mass $m$:
\begin{equation}
 S[Q]
  = \frac{1}{4} \mathop{\mathrm{Str}} \left[
       -\frac{\sigma_{xx}}{2} (\nabla Q)^2
       +\biggl( \sigma_{xy} + \frac{1}{2} \biggr)
         Q \nabla_x Q \nabla_y Q
    \right].
\label{final-action}
\end{equation}
The topological term is equal to $i\theta N[Q]$, with the angle $\theta = 2 \pi
\sigma_{xy} + \pi$. In graphene the mass $m$ is absent, so that
$\sigma_{xy}=0$. Thus, the topological angle is $\theta = \pi$. The theory
(\ref{final-action}) is then exactly on the critical line of the quantum Hall
transition \cite{Pruisken}, in agreement with the arguments of
Ref.~\cite{Ludwig}.  Thus the graphene with a generic (TR breaking)
long-range disorder  is driven into the 
quantum Hall critical point, with the
conductivity $4\sigma^*_U$ (the factor 4 accounts for the spin and valley
degeneracy). The value $\sigma^*_U$ is known to be in the range $\sigma^*_U
\simeq 0.5 - 0.6$ from numerical simulations \cite{Quantum-Hall_critical}. A
schematic scaling function in this case is shown in Fig.\ \ref{Fig:beta}a.
While formally this result holds for any energy $\varepsilon$, in reality it
only works near half filling (where the bare conductivity is $\sim e^2/h$); for
other $\varepsilon$ the quantum Hall critical point would only be reached for
unrealistic temperatures and system sizes.

Let us now turn to the case of preserved TR invariance, describing in
particular charged impurities. The system belongs then to the symplectic
symmetry class AII. The derivation of the $\sigma$-model starts with the
doubling of $\psi$ variables accounting for the TR symmetry \cite{Efetov}. Then
$Q$ is $8\times 8$ matrix obeying an additional constraint of charge
conjugation $Q = \bar Q$. The real part of the action is calculated in the same
way as in the unitary case, yielding Eq.\ (\ref{action_real}) with an
additional factor $1/2$.

Since the partition function of the symplectic model is real, the imaginary
part of the action $S_2[Q]$ can take one of the two possible values, $0$ or
$\pi$. The discreteness of $S_2[Q]$ suggests that it again should be
proportional to a topological invariant on the $\sigma$-model manifold. A
non-trivial topology may arise only in the compact (fermion) sector of $Q$. The
corresponding target space is $\mathcal{M}_F = O(4n) / O(2n) \times O(2n)$,
where $n$ is the number of fermion species. While for the conventional
(``minimal'') $\sigma$-model $n = 1$, larger values will arise if one considers
higher-order products of Green functions. The topological invariant takes
values from the homotopy group \cite{ViroFuchs}
\begin{equation}
 \pi_2 \bigl[ \left. \mathcal{M}_F\right|_{n=1} \bigr]
  = \mathbb{Z} \times \mathbb{Z}, \qquad
 \pi_2 \bigl[ \left. \mathcal{M}_F\right|_{n\geq2} \bigr]
  = \mathbb{Z}_2.
 \label{pi2}
\end{equation}
The homotopy group in the case $n = 1$ is richer than for $n \geq 2$.
Nevertheless, $S_2[Q]$ may take only two non-equivalent values. Hence only a
certain $\mathbb{Z}_2$ subgroup \cite{Fendley} of the whole $\mathbb{Z} \times
\mathbb{Z}$ comes into play as expected: the phase diagram of the theory should
not depend on $n$.

%%%%%%%%%%%%%%%%%%%%%%%%%%%%%%%%%%%%%%%%%%%%%%%%%%%%%%%%%%%%%%%%%%%%%%
%%%%%%%%%%%%%%%%%%%%%%%%%%%%%%%%%%%%%%%%%%%%%%%%%%%%%%%%%%%%%%%%%%%%%%

To demonstrate the emerging topology explicitly and to calculate the
topological invariant, let us analyze the case $n = 1$ in more detail. The
generators of the compact sector are Hermitian skew-symmetric $4 \times 4$
matrices anticommuting with $\Lambda \equiv \rho_3$: $\rho_1 \tau_2$, $\rho_2
\tau_0$, $\rho_2 \tau_1$, and $\rho_2 \tau_3$. Here  $\rho_i$ and $\tau_i$ are
Pauli matrices in RA and TR space respectively. These generators split into two
mutually commuting pairs, each generating a 2-sphere (``diffuson'' and
``Cooperon'' sphere). Simultaneous inversion of both spheres leaves $Q$ intact.
Hence the compact sector of the model is the manifold $(S_2 \times S_2) /
\mathbb{Z}_2$. Thus two topological invariants, $N_{1,2}[Q]$, counting the
covering of each sphere, emerge in accordance with Eq.\ (\ref{pi2}). The most
general topological term is $i S_2 = i\theta_1 N_1 + i\theta_2 N_2$. Due to the
TR symmetry, the action is invariant under interchanging the diffuson and
Cooperon spheres, which yields $\theta_1 = \theta_2 \equiv \theta$ where
$\theta$ is either $0$ or $\pi$. The explicit expression for $n=1$ topological
term can be written using $\mathbf{u} = T \nabla T^{-1}$
\begin{equation*}
 i S_2[Q]
  = \frac{\epsilon_{\alpha\beta}}{8} \mathop{\mathrm{Str}} \bigl[
      (\Lambda \pm 1) \tau_2 u_\alpha u_\beta
    \bigr]
  \equiv i\pi(N_1[Q] + N_2[Q]),
\end{equation*}
yielding $\theta = \pi$. The sign ambiguity here does not affect any
observables. If the TR symmetry is broken, the Cooperon modes are frozen and
the manifold is reduced to a single diffuson sphere with $i S_2[Q] = i\pi
N_1[Q]$.

The ordinary symplectic theory with no topological term exhibits a
metal-insulator transition at $\sigma^*_{Sp} \approx 1.4$ \cite{Schweitzer}. If
the conductivity is smaller than this critical value the localization drives
the system into insulating state, while in the metallic phase, $\sigma >
\sigma^*_{Sp}$, antilocalization occurs. Using the analogy with the quantum
Hall transition in the unitary class, we argue that the topological term with
$\theta=\pi$ suppresses localization effects when the conductivity is small,
leading to appearance of a new attracting fixed point at $\sigma^{**}_{Sp}$.
The position of the metal-insulator transition, $\sigma^*_{Sp}$, is also
affected by the topological term. However, we believe that its change is
negligible: the instanton correction to the scaling function at large
conductivity is exponentially small \cite{Pruisken}, and the value of the
exponential factor $e^{-4 \pi \sigma}$ is still extremely small at $\sigma =
\sigma^*_{Sp}$. A plausible scaling of the conductivity in the symplectic case
with $\theta = \pi$ is sketched in Fig.\ \ref{Fig:beta}. The existence and
position of the new critical point can be verified numerically. Recent
simulations of graphene \cite{Nomura06, Beenakker_numerics} indeed demonstrate
the stability of the conductivity in the presence of long-range disorder. Of
course, in reality there will be always a weak inter-valley scattering, which
will establish the localization and lowest $T$, in agreement with \cite{AEA}.
However, the approximate quantum criticality will hold in a parametrically
broad range of $T$.

Finally, we discuss a connection between our findings and recent results on the
quantum spin Hall (QSH) effect in systems of Dirac fermions with spin-orbit
coupling  \cite{KaneMele}, which in the presence of random potential also
belong to the symplectic symmetry class. Such systems were found to possess two
distinct insulating phases, both having a gap in the bulk electron spectrum but
differing by the edge properties. While the normal insulating phase has no edge
states, the spin-Hall insulator is characterized by a pair of mutually
time-reversed spin-polarized edge states penetrating the bulk gap. The
transition between these two phases is driven by Rashba spin-orbit coupling
strength. The existence of the edge states was attributed to certain
$\mathbb{Z}_2$ topological index different from one studied above. This
topological order is robust with respect to disorder, even if the latter mixes
the valleys. The 2D $\sigma$-model is insensitive to the edge properties of the
sample and does not capture the difference between the two insulating phases. A
suitable effective theory is the \emph{one-dimensional} (1D) $\sigma$-model for
the edge states. The corresponding topological index characterizes the homotopy
group $\pi_1(\mathcal{M}_F)$. This group is again $\mathbb{Z}_2$ that enables a
$\theta$-term with $\theta$ equal to $0$ or $\pi$. The topological term with
$\theta = \pi$ is present if the number of channels is odd. Then the
conductivity of the 1D system equals $e^2/h$ in the long-length limit
\cite{Takane}: one conducting channel survives localization. This is what
happens in QSH systems when a pair of edge states is not localized
\cite{KaneMele}. In the presence of disorder, such systems will  possess three
phases: metal, normal insulator,  and QSH insulator. We expect that generically
there will be a transition between the latter two. The critical theory
discussed in the present work (2D symplectic $\sigma$-model with $\theta=\pi$)
should then describe this QSH transition.

In summary, graphene with long-range disorder shows quantum criticality at half
filling. If the effective TR symmetry of the single-valley system is preserved
(e.g. when Coulomb scatterers are the dominant disorder), the relevant theory
is the symplectic $\sigma$-model with topological angle $\theta=\pi$ and the
minimal conductivity takes a universal value $4\sigma^{**}_{Sp}$. If the TR
symmetry is broken (e.g. by effective random magnetic field due to ripples), 
the system falls
into the universality class of the quantum Hall critical point, with another
universal value $\sigma_{xx}=4\sigma^*_U$. We have argued that the symplectic
critical point describes also the QSH transition \cite{KaneMele}.

We thank F.~Evers, Y.~Makhlin, V.~Serganova, and I.~Zakharevich for valuable
discussions. The work was supported by the DFG -- Center for Functional
Nanostructures and by the EUROHORCS/ESF (IVG).


\begin{thebibliography}{99}
\bibitem{footnote_critical}
Strictly speaking, saddle-point approximation requires $\sigma_{xx} \gg 1$,
whereas $\sigma_{xx} \sim 1$ at $\varepsilon = 0$. 
This should not, however, affect the universal critical behavior of 
the theory governed by symmetry and topology of the problem.
\end{thebibliography}
\end{document}